\documentclass{statsoc}

\usepackage[a4paper]{geometry}
\usepackage{graphicx}
\usepackage[textwidth=8em,textsize=small]{todonotes}
\usepackage{amsmath}
\usepackage{natbib}
\usepackage{threeparttable}
\usepackage{booktabs}
\usepackage{makecell}
\usepackage{comment}
\usepackage[hyphens]{url}
\usepackage{soul} 
\usepackage{float}
\usepackage{multirow}

\title[Dose optimization design]{Dose optimization design accounting for unknown patient heterogeneity in cancer clinical trials}
\author[Author 1 {\it et al.}]{Rebecca B. Silva\\}
\email{rs4025@caa.columbia.edu}
\author{Bin Cheng}
\author{Shing M. Lee}
\address{Department of Biostatistics,
Columbia University Mailman School of Public Health,
New York, 
NY, 
United States.}

\begin{document}
\begin{abstract}
Project Optimus, an initiative by the FDA’s Oncology Center of Excellence, seeks to reform the dose-optimization and dose-selection paradigm in oncology. We propose a dose-optimization design that considers plateau efficacy profiles, integrates pharmacokinetic data to inform the exposure-toxicity curve, and accounts for patient characteristics that may contribute to heterogeneity in response.
The dose-optimization design is carried out in two stages. First, a toxicity-driven stage estimates a safe set of doses. Then, a dose-ranging efficacy-driven stage explores the set using response and patient characteristic data, employing Bayesian Sparse Group Selection to understand patient heterogeneity. Between stages, the design integrates pharmacokinetic data and uses futility assessments to identify the target population among the general phase I patient population. An optimal dose is recommended for each identified subpopulation within the target population. The simulation study demonstrates that a model-based approach to identifying the target population can be effective; patient characteristics relating to heterogeneity were identified and different optimal doses were recommended for each identified target subpopulation. 
Most designs that account for patient heterogeneity are intended for trials where heterogeneity is known and pre-defined subpopulations are specified. However, given the limited information at such an early stage, subpopulations should be learned through the design.  
\end{abstract}

\section{Introduction}

The goal of a phase I dose-finding design in cancer clinical trials is to estimate an optimal dose of a cancer therapy that is safe and shows evidence of efficacy. With the advent of many novel non-cytotoxic cancer therapies, there has been an extensive shift from the traditional dose-finding paradigm designed for chemotherapies to a dose-optimization paradigm suitable for newer oncology therapies such as molecularly targeted therapies (MTAs) and immunotherapies. Project Optimus, initiated by the FDA’s Oncology Center of Excellence, is an ongoing initiative to reform the dose optimization and dose selection paradigm in oncology \citep{fouriezirkelbachImprovingDoseOptimizationProcesses2022, friendsofcancerresearchwhitepaperOptimizingDosingOncology}. The initiative challenges the misconception that higher doses lead to more efficacy and proposes using safety information that integrates PK and PD data and assessing at least two doses in a randomized manner to understand the risk-benefit trade-off across doses. The initiative also emphasizes the importance of considering patient heterogeneity and understanding the dose-response and dose-toxicity curves. 

Many of the criticisms of dose-finding designs in Project Optimus have been addressed, but generally publications have focused on one or two of the issues. 
The consideration of using both toxicity and efficacy outcomes to find an optimal dose was addressed initially through the phase I/II paradigm  \citep{thallBayesianDesignsPhase2016}. Some designs model toxicity and efficacy jointly and consider a risk-benefit tradeoff \citep{thallDoseFindingBasedEfficacy2004, yuanBayesianDoseFinding2009} while others model toxicity and efficacy separately in sequential stages and use criteria for safety and efficacy to find the optimal dose \citep{panPhaseIISeamless2014, wagesSeamlessPhaseII2015, yanGeneralizationTimetoeventContinual2019, guoBayesianPhaseII2022a}. Under this paradigm, adaptive randomization among doses and approaches for accounting for non-monotone efficacy profiles and late-onset of endpoints have been proposed. These designs have been extended to incorporate patient covariates known to differentiate the patient population \citep{thallPatientspecificDoseFinding2008, guoBIPSEBiomarkerbasedPhase2022} or account for known subgroups \citep{wages_phase_2015}. Other methods have been proposed to leave more flexibility in subpopulation specification either through allowing adaptive clustering of predefined subgroups \citep{lee_precision_2021, curtisSubgroupspecificDoseFinding2022} or accounting for multiple patient covariates through utilizing dimension reduction methods, such as canonical partial least squares \citep{guo2017}, the Bayesian Lasso \citep{kakuraiDoseIndividualizationVariable2019}, or a latent subgroup membership model \citep{guoBayesianPhaseII2023}. 
One design has been proposed to address unknown patient heterogeneity and recommend subgroup-specific doses, but this was proposed solely in the context of toxicity \citep{silva_papertox}.  
A major limitation of these methods is that they assume the same population for phase I and II and in most cases, the target population, otherwise known as the phase II eligible population, may not be known before the therapy is tested in patients. Strict enrollment for an assumed target population could lead to missing efficacy signals in unexpected populations \citep{paolettiPhaseIITrial2018}.  Attempts to address this have used dose-expansion cohorts, such as using hypothesis testing to sequentially monitor efficacy in specific patient subgroups\citep{iasonosSequentialMonitoringPhase2017b}  or using randomization and backfill in the dose-expansion phase to assign patients to different doses besides the MRD \citep{dehbiControlledBackfillOncology2021, dehbiControlledAmplificationOncology2023a}.

With the advent of dose optimization, more designs employing both efficacy and toxicity have been proposed that apply novel techniques to compare doses such as incorporating PK \citep{zhangModelBasedTrialDesign}, immune response \citep{ guoDROIDDoserangingApproach}, and long-term endpoints \citep{ thallGeneralizedPhaseIII2023}, utilizing Bayesian dynamic learning \citep{qiuBayesianDynamicModelBased}, or enhancing model-assisted designs with dose optimization approaches \citep{takedaBayesianOptimalInterval2023, dangeloUMETUtilityBasedDosea}. These designs, however, do not focus on the potential heterogeneity in the patient population and the need to identify different doses for heterogeneous subpopulations.

The proposed dose-optimization design aims to maximize the data from dose-finding trials including toxicity, efficacy, PK, and late-onset outcomes while incorporating non-monotone efficacy profiles and addressing unknown heterogeneity in the patient population by identifying potential subpopulations and if needed, differing optimal doses. 
The  design consists of a Dose-escalation and a Dose-ranging stage.
In the Dose-escalation stage, patients are assessed as in a first-in-human study and dose-assignment follows sequentially based on the toxicity information obtained, using a model-based dose-escalation method that accounts for delayed outcomes. In this stage, we aim to identify a set of safe doses which we want to explore further. The set of safe doses is adjusted based on PK information, an integral part of Project Optimus. 
Additionally, before enrolling more patients, we assess futility to evaluate lack of activity in identified subpopulations to estimate the target population and expand the population demonstrating activity. 
In the Dose-ranging stage, doses are explored to optimize efficacy and learn about patient heterogeneity, through an Adaptive Randomization and Optimization phase using Bayesian variable selection. Another futility assessment is conducted to identify the target population. 
Finally, if patient heterogeneity is identified, an optimal biological dose (OBD) is recommended for each identified subpopulation within the target population.

The design is unique in its approach to learning about patient heterogeneity. It not only allows for the optimal dose to differ among patients, but at the same time does not assume certain patient characteristics are influential or subgroups are known.

The rest of the paper is outlined as follows. In Section~\ref{mot_ex}, we describe the motivating example. The dose-optimization design is detailed in Section~\ref{methods}. In Section~\ref{sim}, we present a simulation study based on the redesigned entrectinib trial, and compare the performance of the proposed design to the naive approach of not accounting for any patient heterogeneity. Finally, we discuss the significance and limitations of our proposed design in Section~\ref{discuss}.

\section{Motivating Example}
\label{mot_ex}
The proposed dose-optimization design is motivated by two phase I trials conducted to evaluate the safety and antitumor activity of entrectinib, a selective inhibitor of the tyrosine kinases TRK A/B/C, ROS1, and ALK in patients with locally advanced or metastatic cancer with confirmed NTRK 1/2/3, ROS1, or ALK molecular alterations \citep{drilonSafetyAntitumorActivity2017}. The two phase I trials, study STARTRK-1 ($n = 65$) and study ALKA-372-001 ($n = 54$), assigned patients to body surface area based doses of 100, 200, and 400mg/$\text{m}^2$ and fixed doses of 600mg and 800mg using the traditional $3+3$ dose-escalation design. 

\begin{figure}
\centering
\includegraphics[width=.5\textwidth]{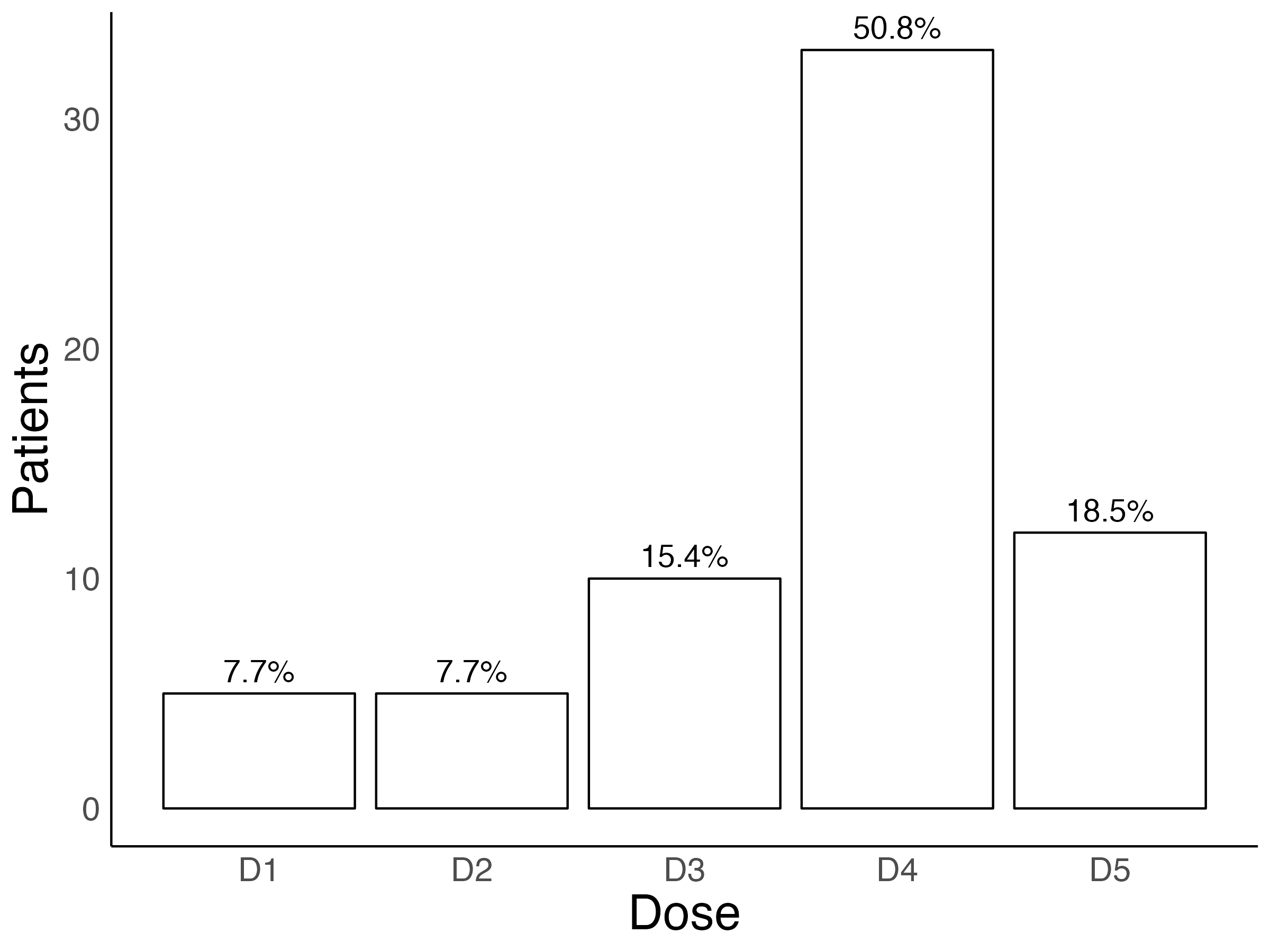}
	\caption{Entrectinib dosing in the STARTRK-1 phase 1 trial: number and percent of patients assigned to each dose. Doses are 100mg/m$^2$, 200mg/m$^2$, 400mg/m$^2$, 600mg and 800mg. MTD and Recommended Dose (RD) = 600mg.}
	\label{entrec_ex}
\end{figure}

Toxicity was indicated by the occurrence of a first-cycle dose-limiting toxicity (DLT) and efficacy was assessed using the objective response rate (ORR). Dose level 4, 600mg, was found to be the fixed-dose MTD at which dose additional patients were added and the recommended phase II dose (RP2D) was identified.
The phase II-eligible population was identified as the subpopulation of tyrosine kinase inhibitor (TKI) treatment-naive patients with a gene fusion involving NTRK 1/2/3, ROS1, or ALK, after analyzing data from the STARTRK-1 and ALKA-372-001 trials. Among the identified phase II-eligible population, a different ORR was found between gene type (NTRK 1/2/3, ROS1, and ALK) at the MTD, however, no exploration around the MTD seems to have been formally conducted, leaving uncertainty about the efficacy at other doses.  Figure~\ref{entrec_ex} shows the distribution of patients among the five doses in the STARTRK-1 trial, where about 50\% of the trial population was assigned to dose level 4 (600mg).
While analysis on PK data was conducted, it is unclear how results informed the RP2D since no formal approach was used. In our design, we propose using a PK-adjusted MTD to identify the upper limit of the acceptable doses that should be explored further. Finally, enrolled patients were heterogeneous with respect to the type of molecular alteration, the location of molecular alteration (NTRK 1/2/3, ROS1, or ALK genes), and other patient characteristics such as prior treatments, however no characteristics were incorporated into dose-finding.  Since many of these patient characteristics are potential predictive markers of the target population and optimal biological dose, a dose-optimization design could have more efficiently learned about target subset populations and accounted for possible heterogeneity in finding the optimal dose.

\section{Methods}
\label{methods}

Consider a total of $J$ doses levels, $d_1, \ldots, d_J$, that will be evaluated in the dose-optimization trial. Using the entrectinib phase I trial as a motivating example, we consider a binary endpoint for toxicity, $Y_T$, and a binary endpoint for efficacy, $Y_E$, where $Y_{Ti} = 1$ and $Y_{Ei} = 1$ indicate a DLT and objective response for patient $i$, respectively. We also account for $M$ patient covariates, $\boldsymbol{z} = (z_1,\ldots,z_M)^T$, which could be responsible for heterogeneity in response. Our objective is to identify the target population that responds to the therapy, identify the subpopulations within the target population that have different optimal doses, and estimate the OBD for each identified subpopulation. We evaluate the two primary endpoints sequentially, starting with toxicity to understand safety, for the first $n_1$ patients, and subsequently assessing patient response and accounting for $M$ patient covariates to learn about the subpopulations that respond to therapy and heterogeneity with respect to optimal dose, for $n_2$ patients. 

This section is organized as follows. The first three subsections describe the toxicity, the PK, and the efficacy models used in the proposed method. Subsection 3.4 presents the schema of the dose-optimization design, and subsection 3.5 provides stepwise instruction on its implementation.

\subsection{Dose-toxicity model}
\label{dose-tox-model}

Our objective in assessing toxicity is to identify a set of acceptable doses, $\mathcal{A}_d$, which should be explored further for efficacy. This is a common approach in many sequential phase I/II designs.
In order to characterize $\mathcal{A}_d$, we estimate a toxicity-based MTD.  
To model the dose-toxicity curve, we use the Bayesian Time-to-Event (TITE) CRM, proposed by \citet{cheungSequentialDesignsPhase2000}. 
Let $X_i = x_i$ be the dose assigned to patient $i$, and let the probability of toxicity at $d_j$ be $\pi_T(p_j, a) = P(Y_{Ti} = 1|X_i = d_j)$, $j =1,\ldots, J$. We use a power working for toxicity, $\pi_T(p_j, a) = p_j^{\exp(a)}, j = 1,\ldots ,J$, where $0<p_1<\cdots <p_J<1$ are the prior probabilities of toxicity at the $J$  doses, otherwise known as the dose skeleton. The traditional TITE-CRM accounts for both complete and partially observed patients, and thus can account for longer-term toxicities while not delaying patient recruitment. 
The dose assigned to the next patient ($n+1$) is defined as
$x_{n+1} = \arg \min_{x \in \{d_1,\ldots,d_J\}} | \widehat{\pi}(x; \hat{a}_n) - p_T |$, where $p_T$ is the target toxicity rate. When $n = n_1$ patients have been assigned a dose, then $x_{n+1}$ is the estimated toxicity-based MTD, denoted $\widehat{{\rm MTD}}_{{\rm TOX}}$. We estimate $\mathcal{A}_d$ as the set of safe doses as all doses up to and including the $\boldsymbol{\text{MTD}_{\text{TOX}}}$,
\begin{equation}
   \widehat{\mathcal{A}_d} = \{ d_j : d_j \le \widehat{{\rm MTD}}_{{\rm TOX}} \}. 
   \label{safeset}
\end{equation}

\subsection{Dose-PK model}

To better understand the dose-concentration relationship, we also estimate a PK-defined MTD, $\boldsymbol{\text{MTD}_{\text{PK}}}$, based on the area under the concentration curve (AUC), and use estimated $\boldsymbol{\text{MTD}_{\text{PK}}}$ to update $\widehat{\mathcal{A}_d}$, the estimated set of acceptable doses. We assume that patient $i$ experiences a toxicity if their AUC exceeds a certain threshold and assume that for the prespecified $n_1$ patients in which toxicity was evaluated, PK-sampling is done and AUC is analyzed. Following \citet{[ursino2017]}, we use a modified version of a hierarchical PK-toxicity model proposed by \citet{pattersonNovelBayesianDecision1999} and \citet{whiteheadEasytoimplementBayesianMethods2001}, to model PK to dose using a linear regression model, 

\begin{equation}
v_i|(\boldsymbol{\beta}, D_{[i]}=D_j) \sim N(\beta_0 + \beta_1 \log(D_j),  \sigma^2), 
\end{equation}
where $D_{[i]}$ is the dosage assigned to patient $i$, and $D_j$ is the dosage at $d_j$,  $v_i = \log(AUC_i)$ and $\boldsymbol{\beta} = (\beta_0, \beta_1)$ with prior distributions $\boldsymbol{\beta}|\sigma \sim N_2(\boldsymbol{m}, \sigma^2 \boldsymbol{G})$ and $\sigma \sim {\rm Beta}(a,b)$. In view of the relationship $AUC_i = D_{[i]}/\text{CL}_i$, $\sigma$ in (2) reflects the inter-individual variation in drug clearance (CL) and constants $\boldsymbol{m}, \boldsymbol{G}, a$, and $b$ reflect prior knowledge about the assumed linear relationship between $v_i$ and $D_{i}$. The logarithm of dose and logarithm of AUC was found to best describe a linear relationship in simulations. Other PK measures that could also be used include the maximum observed plasma concentration or time of maximum concentration \citep{[ursino2017]}.

The probability of toxicity at $d_j$ is estimated as the probability that AUC exceeds a known threshold, $L$, given as $P(v_i > \log(L)|{\boldsymbol{\beta}}, D_j)$. $\boldsymbol{\widehat{\text{MTD}}_{\text{PK}}}$ is defined  as the dose level $D^*$ such that $P(v_i > \log(L)|{\boldsymbol{\beta}}, D^*)$ is closest to the target toxicity rate, $p_T$. We define the adjusted estimated MTD, $\widehat{\text{MTD}}^*$ as 
\begin{equation}
\widehat{\text{MTD}}^* = \min\{\widehat{\text{MTD}}_{\text{TOX}}, \widehat{\text{MTD}}_{\text{PK}}\},
\end{equation} 
following the PKCRM approach described in \citet{[ursino2017]}, and update the set of safe doses, $\widehat{\mathcal{A}^*_d}$, as, 

\begin{equation}
\widehat{\mathcal{A}^*_d} = \{d_j: d_j\le \widehat{\text{MTD}}^*\}.
\label{adjustedsafeset}
\end{equation}

\subsection{Dose-efficacy model} 

Let $Y_E$ be the binary response outcome and assume we have a potentially large collection of patient covariates, $\boldsymbol{z} = (z_1,\ldots,z_M)^T$, which could be responsible for heterogeneity in efficacy. Our objective in evaluating efficacy is to understand the dose-response relationship, considering that $\boldsymbol{z}$ may contribute to the relationship and lead to different dose-response curves across the patient population. We use the estimation of the dose-response relationship to identify the target population and potential subpopulations within the target population, and estimate the optimal dose for each respective subpopulation. 

We model $\pi_E(p_j, \boldsymbol{z}_i, \boldsymbol{\theta}) = P(Y_{Ei} = 1|X_i = d_j, \boldsymbol{z}_i)$ as, 

\begin{equation}
\pi_E(p_j, \boldsymbol{z}_i, \boldsymbol{\theta}) = \exp\{-\exp\{\alpha_0\}\} (1- \exp\{-(\exp\{\alpha_1 + \sum_{m = 1}^M \gamma_m z_{mi}\}p_j\}),
\label{efficacy_model}
\end{equation}
where $\boldsymbol{\theta} = (\alpha_0, \alpha_1, \gamma_1,\ldots, \gamma_M)^T$. Equation (\ref{efficacy_model}) models a dose-efficacy plateau curve such that efficacy is a non-decreasing function of dose where $\alpha_1$ accounts for the dose effect, $\exp\{\sum_{m = 1}^M \gamma_m z_{mi}\}$ represents the shift in efficacy given the patient covariate vector $\boldsymbol{z}_i$, and $\exp\{-\exp\{\alpha_0\}\}$ represents some plateau limit between 0 and 1. 

\begin{figure}[!ht]
\vspace*{\floatsep}
\centering
\includegraphics[width=0.7\textwidth]{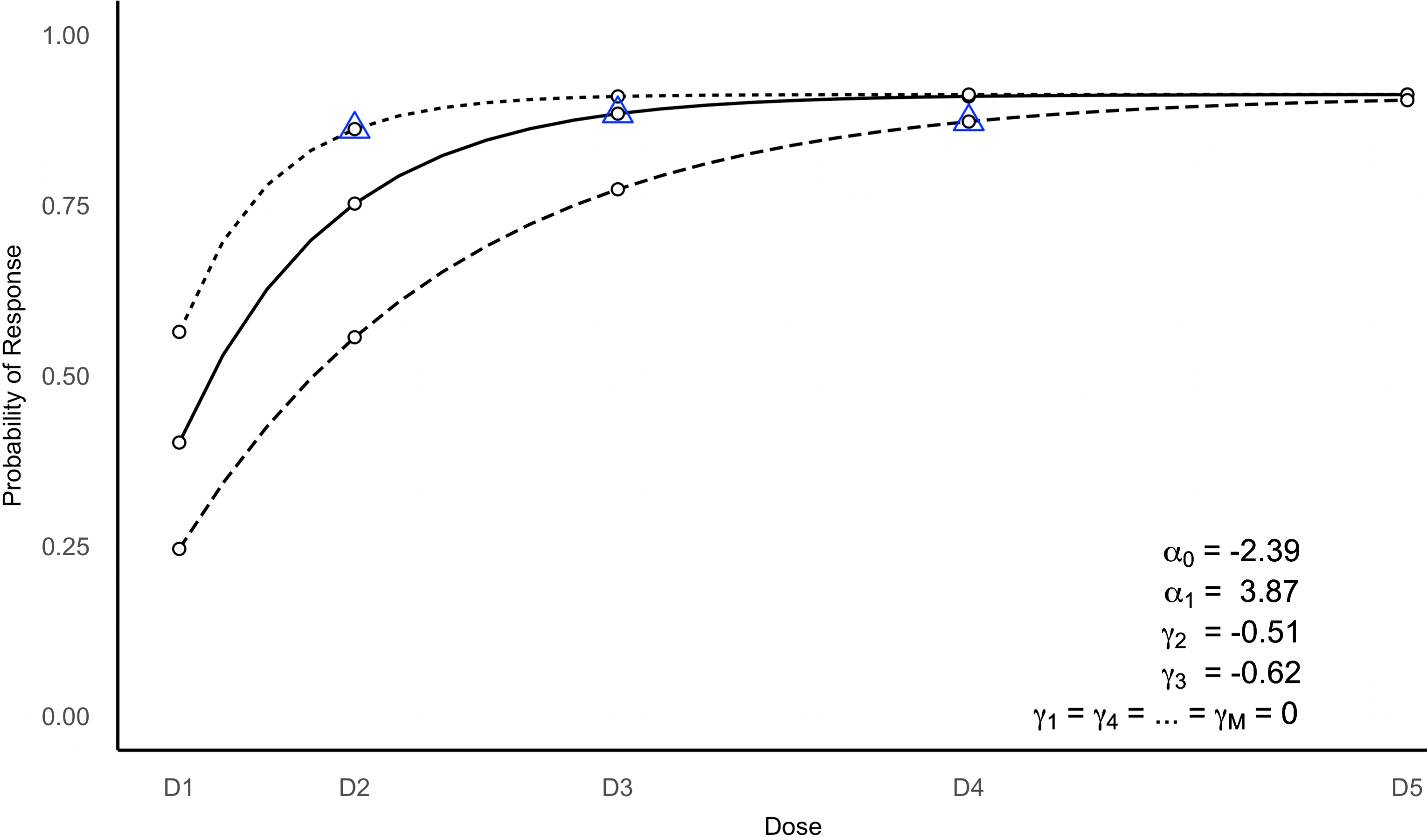}
\caption{Example of the dose-efficacy model with three Optimal Biological Doses (OBD), denoted with a triangle, when $z_2$ and $z_3$ are the only patient covariates associated with efficacy.}
\label{eff_model_ex}
\end{figure}

In addition, we make the sparsity assumption that 
$\sum_{m=1}^M 1_{\gamma_m \neq 0} = K << M$. Specifically, although there could be many potentially relevant patient covariates and we do not have prior knowledge about which affect efficacy, we assume only a small subset of covariates will actually differentiate dose and are clinically relevant to account for in the dose-optimization design. Figure~\ref{eff_model_ex} displays three dose-efficacy curves generated from Equation (\ref{efficacy_model}) when $z_2$ and $z_3$ are the only patient covariates associated with efficacy. When ($z_2 = 0, z_3 = 0$), efficacy plateaus at dose level 2. When ($z_2 = 1, z_3 = 0$) or ($z_2 = 0, z_3 = 1$), efficacy plateaus at dose level 3, and when ($z_2 = 1, z_3 = 1$), efficacy plateaus at dose level 4. Thus, for the target population, there are three subpopulations with different optimal doses.

At such an early phase of drug evaluation, we want to bring forward and identify covariates that signal a strong enough difference in dose that subgroups could be identified by them. The sparsity constraint allows us to bring forward only a few subgroups for realistic clinical applications. Unlike other existing methods, we do not assume we know beforehand these covariates or subgroups going into the trial and want to account for as many potentially influential covariates as possible. 

Given Equation (\ref{efficacy_model}) and the sparsity assumption, we want to identify the subset of $K$ covariates that are responsible for heterogeneity in OBD. 
\citet{oharaReviewBayesianVariable2009} give an overview and comparison of many Bayesian variable selection methods, all of which use some variation of a “spike and slab” prior. \citet{jreichReviewBayesianSelection2022} give an overview of Bayesian selection methods with grouped covariates which extend standard Bayesian variable selection. In application, categorical variables are most relevant here since they can account for binary variables, categorical variables with more than two levels, and categorized continuous variables. For example, PD-L1 expression level in a tumor, which has been shown to be a predictor of response \citep{guoSPIRITSeamlessPhase2018}, can be categorized as binary, however, in a clinical setting it may be more informative to categorize expression into three groups based on Tumor Proportion Score (TPS) \citep{linPrognosticSignificancePDL12017a}.
To identify the influential categorical variables and levels of each variable, we use a modification of the Stochastic Search Variable Selection (SSVS) method called the Bayesian Sparse Group Selection (BSGS) proposed by \citet{chenBayesianSparseGroup2016}. The modification allows for inclusion of categorical variables with more than two levels and the identification of influential levels within each variable by specifying a hierarchical prior for each level within the categorical variable.

Consider $H$ categorical patient characteristics where each characteristic $h$, has $C_h$ levels, $h = 1,\ldots, H$. Then, there are $\sum_{h = 1}^H (C_h-1) = M$ total patient covariates besides dose in the efficacy model, after excluding the reference level of each categorical covariate. The efficacy model from Equation (\ref{efficacy_model}) can be rewritten as, 
\begin{equation}
\pi_E(p_j, \boldsymbol{z}_i, \boldsymbol{\theta}) = \exp\{-\exp\{\alpha_0\}\} (1- \exp\{-(\exp\{\alpha_1 + \sum_{h = 1}^H \sum_{l=1}^{C_h -1} \gamma_{lh} z_{lhi}\}p_j\}).
\label{efficacy_model_cat}
\end{equation}

The BSGS method considers binary indicators for the inclusion of each patient characteristic, $\xi_h, h = 1,\ldots,H$, and each level within the characteristic, $\nu_{lh}, l = 1,\ldots, C_h -1$,  where the level indicators are nested within the characteristic indicators as, 

\begin{equation}
    \xi_h \sim {\rm Ber}(q_h),
\end{equation}
\begin{equation}
    \nu_{lh}|\xi_h \sim (1-\xi_h)\delta_0 + \xi_h {\rm Ber}(q_{lh}),
\end{equation}
where $q_h$ and $q_{lh}$ are the prior probability of inclusion of covariate $h$ and level $l$ in covariate $h$, respectively, for $h = 1,\ldots,H$ and $l = 1,\ldots,C_h -1$, and $\delta_0$ is the Dirac distribution at 0. The inclusion indicator of the regression coefficient corresponding to the $l$th level of category $h$ is then $\eta_{lh} = \nu_{lh}\xi_h$. 

The prior of each regression parameter, $\gamma_{lh}$, is conditional on the inclusion parameter, $\eta_{lh}$, and can be written as the mixture of a point mass at zero, $\delta_0$, and a Gaussian distribution,  
\begin{equation}
    \gamma_{lh}|\nu_{lh}\xi_h \sim (1- \nu_{lh}\xi_h) \delta_0 + \nu_{lh}\xi_h N(0, \tau_{lh}^2). 
    \label{gaussian_mixture}
\end{equation}

That is, if $\nu_{lh} = \xi_h = 1$, the prior distribution is a wide distribution around 0 with variance $\tau_{lh}^2$, the ``slab", and if $\nu_{lh} = 0$ or $\xi_h = 0$, the prior distribution is a point mass at 0, the ``spike". 

The parameters $q_h$, $q_{lh}$, and $\tau_{lh}$, $h = 1,\ldots, H$, $l = 1,\ldots, C_h - 1$ must be tuned for desired model performance. We use $q_h = 0.5$ and $q_{lh} = 0.5$ for $h = 1, \ldots, H$, $l = 1,\ldots, C_h - 1$ based on findings by \citet{jreichReviewBayesianSelection2022} that for small sample sizes, fixing $q_h$ and $q_{lh}$ demonstrated more robust correct selection of covariates compared to Uniform or Beta distributions. We chose $\tau_{lh}$ to represent a reasonable range for $\gamma_m$ and further tuned $\tau_{lh}$ in simulation studies for reasonable variable selection. Inclusion of each covariate and their respective levels were based on the mean posterior inclusion probability of $\eta_{lh}$ satisfying some threshold, $\Psi_E$. Specifically, level $l$ of covariate $h$ was included if $P(\eta_{lh} = 1)> \Psi_E$. 

Furthermore, we modify the prior distribution of $\gamma_{lh}$ by imposing a truncated Gaussian distribution instead of a Gaussian distribution in the mixture model when the direction of association is known for all levels of category $h$. If less is known about a covariate, but investigators still wish it to be in the model, a Gaussian mixture model as in Equation (\ref{gaussian_mixture}) is used. This allows flexibility in specifying different mixture priors for each patient covariate. For example, PD-L1 positive patients have shown improved response compared to PD-L1 negative patients across many cancer types but both expression types may benefit from a therapy, so the direction of expression type adds information to the model.

Just as our dose-toxicity model with the TITE-CRM accounts for partially observed toxicity outcomes, we account for partially observed efficacy outcomes using a weighted binomial likelihood. Given efficacy data for $n$ patients, \[ D_{En} = \{(Y_{E1}, w_{E1}, x_1),\ldots,(Y_{En}, w_{En}, x_n)\},\] the weighted likelihood is
\begin{equation}
L(\boldsymbol{\theta}|D_{En}) = \prod_{i=1}^n [w_{Ei} \pi_E(x_i, \boldsymbol{z}_i, \boldsymbol{\theta}) ]^{Y_{Ei}} [1-w_{Ei} \pi_E(x_i, \boldsymbol{z}_i, \boldsymbol{\theta}) ]^{1-Y_{Ei}},
\end{equation}
where $w_{Ei}$ is the efficacy weight representing the proportion of efficacy follow-up time of the $i$th patient: $w_{Ei} = \min(t_i/T_E, 1)$ when $Y_{Ei} = 0$ and $w_{Ei} = 1$ when $Y_{Ei} = 1$, where $T_E$ is the efficacy observation period and $t_{i}$ is the length of time patient $i$ has been followed. 

The posterior distribution of $\boldsymbol{\theta} = (\alpha_0, \alpha_1, \gamma_{1,1}, \ldots, \gamma_{C_H -1, H})$ is then, 

\begin{equation}
    f(\boldsymbol{\theta}|D_{En}) \propto L(\boldsymbol{\theta}|D_{En}) f(\alpha_0) f(\alpha_1) \prod_{h = 1}^H \left(\prod_{l = 1}^{C_h -1} f(\gamma_{lh}|\nu_{lh}\xi_h) f(\nu_{lh}|\xi_h)\right)f(\xi_h), 
\end{equation}
where $f(\cdot)$ denotes the prior distribution of the parameters in $\boldsymbol{\theta}$. We use the `R2jags' package in R to implement Markov chain Monte Carlo (MCMC) simulations and sample the posterior distribution to compute the posterior estimates of $\boldsymbol{\theta}$. Given $\widehat{\boldsymbol{\theta}}_n$, we obtain the estimated probability of efficacy for the $(n+1)$-th patient, $\pi_E(p_j, \boldsymbol{z}_{n+1}, \widehat{\boldsymbol{\theta}}_n)$ from Equation~(\ref{efficacy_model_cat}). 

Depending on the stage of the dose-optimization trial, we assign a dose to the $(n+1)$-th patient using different approaches that depend on $\pi_E(p_j, \boldsymbol{z}_{n+1}, \widehat{\boldsymbol{\theta}}_n)$, described in Section~\ref{steps}. The approaches used to identify the target population and the OBD, which use the estimated probability of efficacy are also detailed in Section~\ref{steps}.

\subsection{Dose-optimization design}

\begin{figure}[!ht]
\vspace*{\floatsep}
\centering
\includegraphics[width=\textwidth]{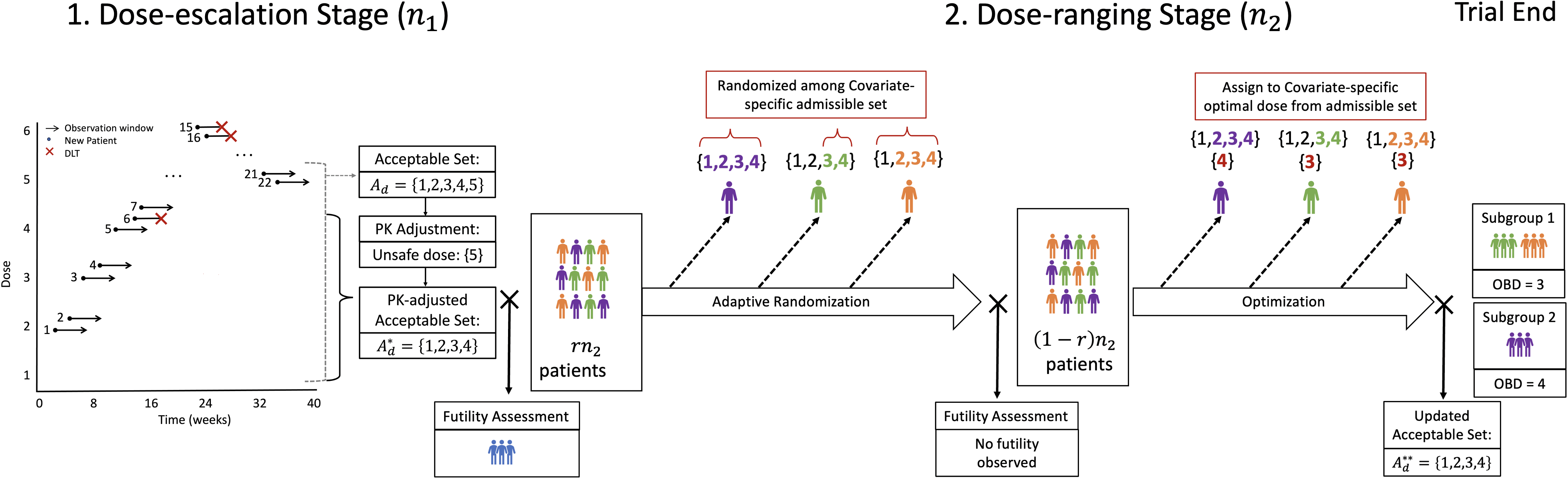}
\caption{Schema of Dose-Optimization Design. OBD, Optimal Biological Dose. $r \in (0,1)$.}
\label{schema_figure}
\end{figure}

The proposed dose-optimization design uses model-based approaches to define a set of acceptable doses, identify the target population, and find the patient characteristics that affect efficacy, thus determining the subpopulation-specific OBD. A schema of the proposed design is shown in Figure~\ref{schema_figure}. In the initial toxicity-focused stage, the Dose-escalation stage, an initial acceptable set of doses is identified after sequential dose-escalation of $n_1$ patients. This is done through using the TITE-CRM which estimates a DLT-based MTD after $n_1$ patients have been observed, discussed in Section~\ref{dose-tox-model}. 
In this initial stage, we assume a realistic phase I population, in which safety can be evaluated for all, but a proportion of patients may not respond to the therapy, as they are not in the true target population. To integrate additional information on safety, we incorporate PK data through fitting a dose-PK model to find the upper limit of safety according to the observed PK data, and adjust the set of acceptable doses using a PK-based MTD, as defined in Definition (\ref{adjustedsafeset}). 
Before more patients are enrolled into the efficacy-focused Dose-ranging stage, a futility assessment is conducted, where we introduce patient characteristics and initial evidence of efficacy is evaluated across the patient population. We use a model-based approach to estimate the target population and discontinue enrollment for any identified subpopulation that is not responding to the therapy. We call this step a futility assessment. 
The Dose-ranging stage follows, in which patient characteristics are accounted for and the probability of response may depend on these characteristics if identified through Bayesian variable selection. Given the estimated probability of efficacy for each patient, before each dose assignment, an admissible set is constructed from doses in the acceptable set at which the probability of response exceeds a lower bound of efficacy. For dose-assignment, the Dose-ranging stage contains an Adaptive Randomization phase, in which patients are randomized to a dose within their respective admissible set, and an Optimization phase, in which patients are assigned their respective optimal dose. Another futility assessment is conducted between the Adaptive Randomization phase and the Optimization phase so that the patients not responding to the therapy are not enrolled further. Lastly, once $N_{\max} = n_1 + n_2$ patients are assigned a dose, and their toxicity and efficacy observation periods have ended, the set of acceptable doses is updated again to include all patients and the OBD for each identified subpopulation is estimated.

\subsection{Dose assignment procedures and steps}
\label{steps}

\textbf{1) Dose-escalation Stage ($n_1$)}

For the first $n_1$ patients, we use the TITE-CRM with the power model to carry out dose-escalation. Here, we only use toxicity data to inform the next dose assignment, although for each patient, the covariate data and efficacy outcome are collected. Similar to most phase I trials, we only use toxicity to assign doses at this stage since PK data is not readily available after each patient. We allow no skipping of dose and before the $n_1$th patient, we assume that the toxicity observation window has been fully observed for each patient. The $\widehat{{\rm MTD}}_{{\rm TOX}}$ is obtained after $n_1$ patients are assigned a dose and the set of acceptable doses, $\widehat{\mathcal{A}}^*_d$, is estimated using Definition~(\ref{safeset}).

At the end of dose-escalation, PK data becomes available, and we use PK data to inform the range of acceptable doses. Using the dose-PK model described previously, we obtain the $\widehat{{\rm MTD}}_{{\rm PK}}$, and adjust the set of acceptable doses, $\widehat{\mathcal{A}}^*_d$, from Definition~(\ref{adjustedsafeset}). 

Lastly, after $n_1$ patients are followed until the end of their observation period, the scaled dose is updated as the posterior probability of toxicity, $\tilde{p}_j = p_j^{\exp(\hat{a}_{n_1})}, j = 1,\ldots,J$, where $\hat{a}_{n_1}$ is the posterior mean.

\begin{figure}[!ht]
\centering
\includegraphics[width=.6\textwidth]{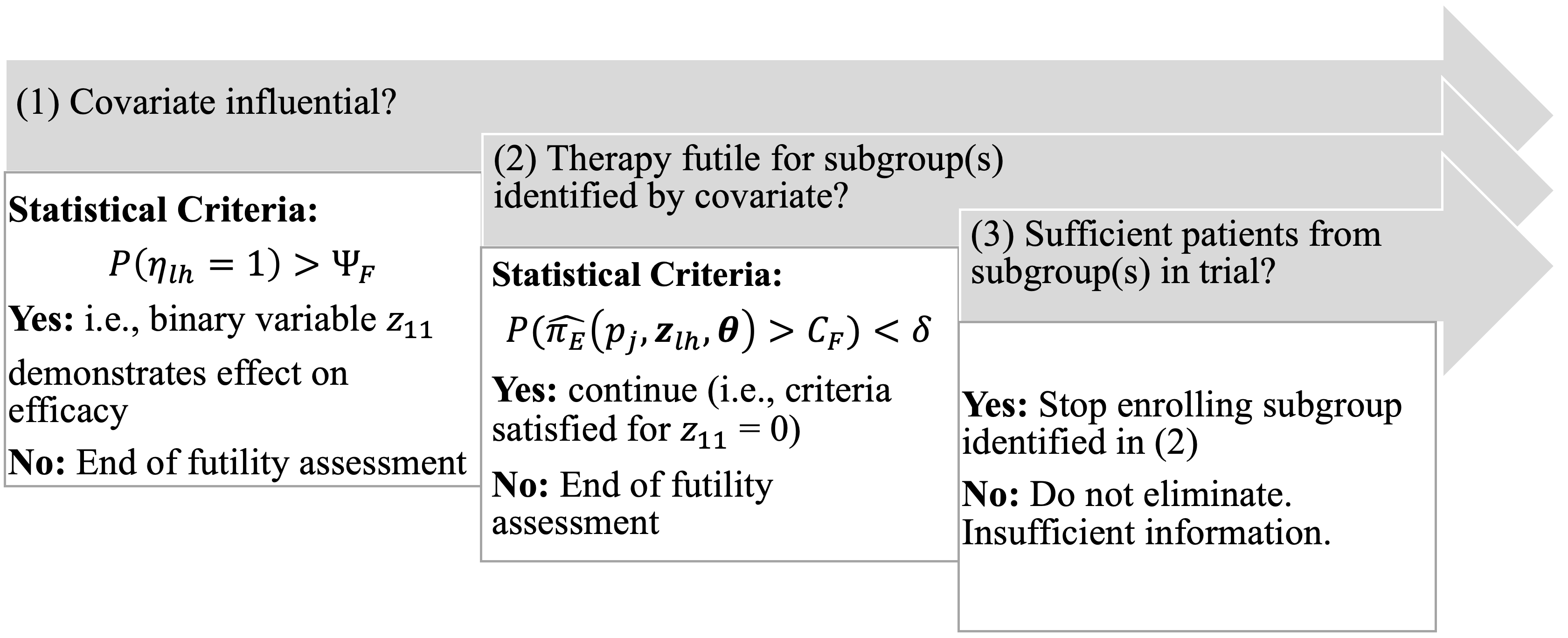}
\caption{Steps in Futility Assessment. The diagram supposes that binary variable $z_{11}$ is found to substantially affect efficacy.}
\label{futility_flow}
\end{figure}

\textbf{Futility Assessment 1:}
Before enrolling more patients into the trial, the first futility assessment is conducted. The futility check provides a model-based approach to determining the target population. The three steps used in each futility assessment are depicted in Figure~\ref{futility_flow}. Given $n_1$ current patients in the trial, we first check if a covariate exists that largely differentiates efficacy. We model efficacy using Equation (\ref{efficacy_model_cat}) with priors specified in Equation (\ref{gaussian_mixture}). Patient covariates are chosen based on their respective posterior inclusion probability exceeding some cutoff, $\Psi_F$, such that $z_{lh}$ is chosen if $P(\eta_{lh} = 1) > \Psi_F$. If more than one covariate is influential, we assess the most influential covariate. If no covariates are influential, we stop the futility assessment. Given a covariate is chosen as influential, we move to step two and assess whether any subgroups identified by the covariate show sufficient evidence that the therapy is futile. 
That is, if some $z_{lh}$ is chosen, we find whether the probability that efficacy exceeds some cutoff, $C_F$, at the estimated highest acceptable dose, $\widehat{\text{MTD}}^*$, is less than some cutoff, $\lambda$,
\begin{equation}
    P(\hat{\pi}_E(p^*_j, z_{lh},\boldsymbol{\hat{\theta}} )> C_F) < \lambda,
\label{futilitycriteria}
\end{equation}
 for $z_{lh} = 0$ and $z_{lh} = 1$ separately, assuming $z_{lh}$ is a binary variable. The efficacy cutoff, $C_F$, should be provided by clinicians as the level of efficacy needed to consider the therapy beneficial for that subpopulation, and the Bayesian threshold, $\lambda$, should be calibrated. Parameter values that have been used are ($C_F = 0.50, \lambda = 0.05$) \citep{curtisSubgroupspecificDoseFinding2022} and ($C_F = 0.30, \lambda = 0.10$) \citep{guoBayesianPhaseII2023} for similar criteria used as stopping rules. If the criterion is satisfied, we move to the last step and determine whether a reasonable number of patients in the subgroup have been assessed to have sufficient evidence of futility. For example, if only $3$ patients have been observed in a subgroup, it might be too early to stop their enrollment given the uncertainty. For example, \citet{curtisSubgroupspecificDoseFinding2022} specify a stopping rule for enrolling a pre-defined subgroup only after the first $6$ patients in the subgroup have been observed, given acceptable doses for the subgroup exist. This number of patients can be guided by simulation studies and the therapy-specific efficacy threshold, $C_F$. After determining whether a sufficient number of patients in the subgroup identified have been observed and criteria (\ref{futilitycriteria}) is satisfied by the subgroup, the therapy is deemed futile for the identified subgroup and the covariate regression effect is removed from the efficacy model. For example, if efficacy for the subgroup identified as $z_{lh} = 0$ is futile, only patients with $z_{lh} = 1$ are enrolled for the rest of the trial. Lastly, if all subgroups identified by the influential covariate are identified as futile, then no target population exists, and the trial terminates.

\textbf{2) Dose-ranging stage:}

Given the set of acceptable doses identified which are both safe and potentially efficacious, we explore the doses through an Adaptive Randomization phase and home in on doses around the estimated subgroup-specific OBD through an Optimization phase. A subset of the $n_2$ patients assigned to the Dose-ranging stage are assigned a dose in each phase, specifically $rn_2$ to the Adaptive Randomization phase and $(1-r)n_2$ to the optimization phase, where $r\in (0,1)$ is a design parameter to be calibrated to optimize optimal dose selection depending on the number of dose levels and maximum number of patients. In \citet{wagesSeamlessPhaseII2015}, two phases were also used and they found through simulation that using half of the total sample size for each phase worked well \citep{wagesSeamlessPhaseII2015}. In both stages, we learn about patient covariates using the Bayesian variable selection method BSGS. 

For each new patient, using all previous data, we model efficacy using Equation (\ref{efficacy_model_cat}) with priors specified for patient covariates such that BSGS selects only informative covariates. Patient covariates are chosen based on their respective posterior inclusion probability exceeding some cutoff, $\Psi_E$, such that $z_{lh}$ is chosen if $P(\eta_{lh} = 1) > \Psi_E$. If $k$ patient covariates are chosen, the posterior estimates are $\widehat{\boldsymbol{\theta}} = (\hat{\alpha}_0, \hat{\alpha}_1, \hat{\gamma}_{(1)}, \ldots, \hat{\gamma}_{(k)})$.  Based on the covariate pattern of the next patient, $\boldsymbol{z}_{n+1}$, the probability of efficacy is estimated at each dose within $\widehat{\mathcal{A}}^*_d$ as, $\hat{\pi}_{E,n+1}(p_j,\boldsymbol{z}_{n+1},\widehat{\boldsymbol{{\theta}}}), d_j \in \widehat{\mathcal{A}}^*_d$. An admissible set for the patient's covariate pattern, $\widehat{\mathcal{A}}^{*}_{\boldsymbol{z}_{n+1}}$, is then constructed from the acceptable set $\widehat{\mathcal{A}}^{*}_d$, excluding any dose that has a probability of efficacy below a cutoff, $\kappa$, and at least $s$ have been treated on the dose, following \citet{rivierePhaseIIDosefinding2018a} who use $\kappa = 0.20$ and $s = 3$. The cutoff $\kappa$ should be set based on the therapy in question.

\textit{Adaptive Randomization phase}:

For the first $r n_2$ patients in the Dose-ranging stage, the new patient is randomized to dose $j$ with probability $R_{j,\boldsymbol{z}_{n+1}}$ proportional to the estimated efficacy: 

\begin{equation}
    R_{j,\boldsymbol{z}_{n+1}} = \frac{\hat{\pi}_{E,n+1}(p_j, \boldsymbol{z}_{n+1}, \boldsymbol{\widehat{\theta}})}{\sum_{d_j \in \widehat{\mathcal{A}}^{*}_{\boldsymbol{z}_{n+1}}} \hat{\pi}_{E,n+1}(p_j, \boldsymbol{z}_{n+1},\boldsymbol{\widehat{\theta}})}.
\end{equation}

The Adaptive Randomization phase allows exploring all admissible doses, as opposed to always assigning patients to the estimated optimal dose, which could lead to getting stuck at a suboptimal dose early in the trial if not enough doses are sufficiently explored. 

\textit{Futility Assessment 2}

A second futility assessment is conducted between the Adaptive Randomization phase and Optimization phase, using the same three steps as Futility Assessment 1.

\textit{Optimization phase}:

In the Optimization phase, we wish to home in on the OBD, estimated as the minimum dose that achieves a probability of efficacy within a certain threshold, $\alpha$, of the maximum efficacy estimated among admissible doses. This is in line with estimating a plateau point as done in \citet{rivierePhaseIIDosefinding2018a} for one target population. After $rn_2$ patients have been assigned a dose, the newly enrolled patient is assigned his respective `optimal' dose, 
\begin{equation}
x_{n+1} = \min\{ d_j: |\hat{\pi}_{E,n+1}(p_j, \boldsymbol{z}_{n+1},\widehat{\boldsymbol{\theta}}) - \max_{d_i \in \widehat{\mathcal{A}}^{*}_{\boldsymbol{z}_{n+1}} }(\hat{\pi}_{E,n+1}(p_i, \boldsymbol{z}_{n+1}, \widehat{\boldsymbol{\theta}}))| < \alpha \}. 
\end{equation}

Instead of fixing $\alpha$, we suggest a calibration-based approach in which $\alpha$ is a decreasing function of $n$, the number of patients currently assigned a dose in Dose-ranging stage, such that as more information from patients is obtained, the threshold becomes more stringent. One such function, modified from \citet{rivierePhaseIIDosefinding2018a} is $\alpha = 0.20(1-0.10 \frac{n}{n_2})$, where $n$ denotes the number of patients currently assigned a dose in the Optimization stage.

\textbf{Identifying the OBD}

After all patients have been fully observed, we use the TITE-CRM to estimate the updated posterior probability of toxicity at each dose and refine the set of acceptable doses, \[ \widehat{\mathcal{A}}^{**}_d = \{d_1,\ldots, \widehat{\text{MTD}}_{n_1+n_2}\}, \]
where $\widehat{\text{MTD}}_{n_1+n_2}$ is the estimated MTD after $n_1 + n_2$ patients are fully observed of their respective toxicity observation period. Since PK data is usually only obtained after the initial dose-escalation stage, we do not use PK to adjust the dose set here.

We fit a final efficacy model as in Equation (\ref{efficacy_model_cat}) with BSGS priors to identify the patient covariates that define different patient subpopulations for the OBD. If $k>0$ covariates, $z_{lh_1},\ldots,z_{lh_k}$, are selected in the final model, let $\boldsymbol{z}_{k^*} = (z_{lh_1},\ldots,z_{lh_k})^T$. Since we are assuming a plateau or positive relationship between efficacy and dose, we want to find the minimum dose that has a high probability of reaching an efficacy that is close to the maximum efficacy achieved by an acceptable dose. Therefore, we define the OBD as the minimum dose at which the posterior probability of response achieving some proportion, $\epsilon$, of the maximum efficacy is greater than some threshold, $\Psi_{\text{OBD}}$. Specifically, the OBD for the subpopulation with covariate pattern $z_{k^*}$ is defined as,

\begin{equation}
    \widehat{{\rm OBD}}_{z_{k^*}} = \{\min_{d_j \in \widehat{\mathcal{A}}^{**}_d}: P(\hat{\pi}_E(p_j,\boldsymbol{z}_{k^*}, \boldsymbol{\widehat{\theta}}) \ge \epsilon \max_{d_i \in \widehat{\mathcal{A}}^{**}_d}\{ \hat{\pi}_E(p_i,\boldsymbol{z}_{k^*}, \boldsymbol{\widehat{\theta}})\}) > \Psi_{\text{OBD}}\},
    \label{obd_criteria}
\end{equation}
where $\hat{\pi}_E(p_j,\boldsymbol{z}_{k^*}, \boldsymbol{\widehat{\theta}}), j = 1,\ldots, J$ is estimated from MCMC chains in which $\boldsymbol{z}_{k^*}$ is selected, ($\eta_{lh_1} = 1,\ldots,\eta_{lh_k} = 1$). Criterion (\ref{obd_criteria}) is used in \citet{guoBayesianPhaseII2023} to identify a subpopulation OBD in which the subpopulation is already identified, however it is also applicable for our proposed design in which variable selection identifies the subpopulation and thus the OBD. Parameters $\epsilon$ and $\Psi_{\text{OBD}}$ should be calibrated over a range of dose-efficacy curves for optimal performance. Lastly, if the estimated efficacy satisfies criteria 
$P(\hat{\pi}_E(\max_{d_j}\{\mathcal{A}^{**}_d\},\boldsymbol{z}_{k^*}, \boldsymbol{\widehat{\theta}}) \ge C_F) < \delta$,
where $C_F$ is the efficacy cutoff and $\delta$ is the Bayesian threshold used in futility Criteria (\ref{futilitycriteria}), then the efficacy is not sufficient and no OBD exists for patients identified by the covariate pattern $z_{k^*}$. We consider this as a final futility assessment, since no OBD implies the therapy is not beneficial for the specific subpopulation identified.

\section{Numerical Studies}
\label{sim}

\subsection{Application to motivating example}

We perform an extensive simulation study motivated by the phase I trial of entrectinib described previously. Through a range of scenarios, including a scenario representing what was observed in the STARTRK trial, we show how the proposed design can identify the target population and different OBDs for subgroups within that population. 

We use the STARTRK trial's binary efficacy and toxicity outcomes and patient covariate information to evaluate doses 100mg/$m^2$, 200mg/$m^2$, 600mg, and 400mg/$m^2$, using a starting dose of 100mg/$m^2$ ($J = 4$). We account for four patient characteristics ($H = 4$): two binary variables indicating prior ROS1/ALK treatment inhibitors and gender, a categorical variable of gene type with three levels, indicating the location of the molecular alteration on NTRK 1/2/3, ROS1, or ALK, and another categorical variable of molecular alteration type with three levels for a gene rearrangement (fusion), amplification, or `other' alteration type. This leads to $M = (2-1)\times 2 + (3-1)\times 2 = 6$ patient regression covariates to consider in the efficacy model, after excluding reference levels. Even though the aim per protocol was to enroll patients with NTRK 1/2/3, ROS1, and ALK molecular alterations, 4 out of the 65 patients enrolled in STARTRK had no known molecular alterations, and we do not include this subset in the trial simulations. We use the distribution of prior treatment, gender, gene location, and molecular alteration from the trial described by \citet{drilonSafetyAntitumorActivity2017}; 34\% of patients had been treated with a prior ROS1/ALK inhibitor, 48\% of patients were female, 18/61 (30\%), 21/61 (34\%), and 22/61 (36\%) of patients had NTRK 1/2/3, ROS1, and ALK molecular alterations, respectively, and 53\%, 16\% and 31\% of patients had a molecular alteration type of gene fusion, molecular amplification, or another mutation, respectively. We impose a truncated Gaussian distribution for prior treatment, assuming the probability of response is lower for those with prior treatment, and gender, assuming males have a higher probability of response, and mutation alteration type, assuming the probability of response increases from `other' (deletions and single point mutations) to `molecular amplification' to `gene fusion', based on assumptions of the target population prior to the phase I trial.

Table 1 
shows the $8$ scenarios we assess, describing different dose-toxicity and dose-response relationships and different target subpopulations within the trial population. 
In Scenarios 1, 5, and 6, we set the probability of efficacy at the true OBD to be the estimated ORR rate identified by \citet{drilonSafetyAntitumorActivity2017} for each gene type. In Scenarios 1, 2, 3, and 4, there is no heterogeneity in OBD, so one dose should be recommended to the target population. In Scenarios 1, 2, 5, and 6, there is heterogeneity in therapeutic benefit and thus the target population is only a subset of the trial population. Specifically, in Scenario 1, three of the four doses are safe and those who do not have a gene rearrangement (fusion) do not respond to the therapy. Thus, the mutation type determines the target population. The OBD is dose level 3 for all gene fusion patients, even though the ORR still differs across gene type. In this scenario, it would be important to identify that  mutation type differentiates the patient population who respond and identify dose level 3 as the OBD for the target population.  Scenario 2 is similar to Scenario 1, except the target population is now TKI-treatment naive patients with a gene fusion, as was the case in the entrectinib trial. Thus, prior treatment and molecular alteration differentiate the trial population with respect to who responds and for the target population, the OBD is dose level 3. In Scenarios 3 and 4, we assume no patient heterogeneity exists, and the target population is the entire patient population enrolled, and one OBD should be recommended for all patients. In Scenario 3, all four doses are acceptable and dose level 2 is the OBD, representing a plateau efficacy profile. In Scenario 4, three of the four doses are acceptable and dose level 3 is the OBD, representing an increasing efficacy profile among acceptable doses. In Scenarios 5, 6, 7, and 8, the target population is heterogeneous and more than one OBD should be recommended. 
Scenario 5 and 6 are similar to Scenario 1 with the same set of acceptable doses and target population, but now the OBD depends on gene type. In Scenario 5, the OBD within the target population depends only on whether a patient harbors gene fusion in genes NTRK 1, 2, or 3, leading to an OBD of 1 while otherwise, patients harboring gene fusion in genes ROS1 or ALK have an OBD of 3. In Scenario 6, the OBD depends on each gene type, leading to an OBD of 1, 2, or 3, respectively, within the target population. In Scenario 7, all four doses are acceptable, and the target population is the entire patient population enrolled, as in Scenario 3, but now the population is heterogeneous with respect to gender, where males have an OBD of dose level 4 and females have an OBD of dose level 2. Lastly, in Scenario 8 the target population is also the entire patient population enrolled, except now only two out of four doses are acceptable and the OBDs are dose level 1 for females and dose level 2 for males.   

We accounted for partially observed follow-up time so that patient accrual is not suspended. Patient accrual followed a Poisson distribution with a rate $= 0.50$ per week over a 4-week period for toxicity ($T_T = 4$) and 8 week observation period for efficacy ($T_E = 8$) per the STARRTRK protocol indicating when DLT and scans for response were evaluated \citep{drilonSafetyAntitumorActivity2017}.  We assumed the time-to-toxicity and time-to-efficacy follow a conditional uniform distribution, meaning the time for each outcome was randomly generated on the interval $(0,T_{i}), i = T,E$ \citep{yanGeneralizationTimetoeventContinual2019}.

We generated the patient’s toxicity and efficacy outcomes, separately, assuming the PK measure of AUC is associated with toxicity and efficacy curves have an increasing or plateau relationship with dose. To generate toxicity, we first simulated PK data. A first-order absorption and elimination linear one compartment model was found to characterize entrectinib well in the STARTRK trial, with parameters of clearance ($CL$), volume of distribution (V), first-order absorption rate constant ($k_a$), and duration of zero-order absorption ($D_1$) \citep{gonzalez-salesPopulationPharmacokineticAnalysis2021}. 
We assumed $CL$ follows a log-normal distribution with a clearance mean of $19.6$L/h and between-patient variability of clearance of $30.8$, $\omega_{CL} = 0.308$, as found by \citet{gonzalez-salesPopulationPharmacokineticAnalysis2021} to get AUC, since $\text{AUC} = \text{dose}/CL$.

To relate PK to toxicity, we assume toxicity occurs if the patient's AUC exceeds a certain threshold, $\tau_L$. Following \citet{[ursino2017]}, for each patient $i$, we generated clearance, $CL_i$, and obtain patient $i$'s AUC at each dose, $D_j$, as $AUC_{ji} = D_j/CL_i$. 
The probability of toxicity can then be modeled as, 

\begin{equation}
    \pi_T(D_j) = \Phi \left( \frac{\log(D_j) - \log(CL_i) -  \log(\tau_L)}{\sqrt{\omega^2_{CL}}}\right),
    \label{pk-lognormal}
\end{equation}
where $\Phi$ is the cumulative function of a standard normal distribution.  We first derive $D_j$ from setting  $\pi_T(D_j) = (0.05, 0.12, 0.25, 0.38)$, where $\text{MTD}_{{\rm PK}}$ is set to dose level 3, and fix $\tau_L$ to $46.31 \text{mg L}$ $\text{h}^{-1}$, found as the highest mean AUC across dose levels in \citet{drilonSafetyAntitumorActivity2017}, and back-solve Equation (\ref{pk-lognormal}).
We subsequently adjusted $\tau_L$ to get different toxicity curves using the obtained dose. Note that we assume that $\tau_L$ is known since PK-adjustment is done only when prior knowledge exists about PK. 

\subsubsection{Trial settings}

In the Dose-escalation stage, we use the TITE-CRM and started at dose level 1 with a target toxicity rate, $p_T$, of 0.25 and cohort size of 1. It has been shown that a target of less than 33\% in the $3+3$ design, as specified in the STARTRK protocol, is most similar to a target rate of 0.25 using the CRM \citep{[silva]}. In the empirical toxicity model, we assume a normal prior of mean of $0$ and variance of $1.34$ on the dose covariate $a$. The dose-toxicity model was calibrated to select a dose that yields between a 19\% and 31\% DLT rate \citep{leeModelCalibrationContinual2009} with a total sample size for the dose-escalation as $n_1 = 24$.

For the Dose-PK model, we used the prior distribution constants as specified in \citet{[ursino2017]}, $a = 1$, $b = 1$, $\boldsymbol{m} = (-\text{log} CL,1)$, $\boldsymbol{G} = \text{diag}(1000, 1000)$ where $CL$ was set as the mean estimated clearance for entrectinib in \citet{gonzalez-salesPopulationPharmacokineticAnalysis2021}. We set $\log(L) = \tau_L$, assuming that the PK model is only fit when there is accurate prior knowledge about the threshold for AUC. 

In the Dose-ranging stage, we assign each dose in cohorts of 1. We set $n_2 =36$ and $n_2 = 56$ for a total trial size, $N_{\max}$, of 60 and 80. In the efficacy model, the priors for the plateau and dose parameters, $\alpha_0$ and $\alpha_1$, were assumed to follow normal distributions with a mean of 0 and variance of 5, since a shift of 2 standard deviations ($2\sqrt{5}$) changes the probability of response by about $0.90$, demonstrating the range of probability of response. For the variance of $\gamma_m, m = 1,\ldots,5$, we set $\tau_{lh} = \tau_h$ as 5, found through tuning in simulations \citep{chenBayesianSparseGroup2016}. Each time the efficacy model was fit, three chains of 2000 MCMC iterations with a burn-in of 1000 MCMC iterations was run in `R2jags'.

Bayesian variable selection was determined using a posterior mean cutoff of $\Psi_E = 0.50$, meaning the covariate was chosen in at least $50$\% of the MCMC models. We set the proportion randomized to the Adaptive Randomization phase equal to the proportion randomized to the Optimization phase, $r = 0.5$. In the Optimization phase, our $\alpha$-function was $\alpha = 0.40(1-0.5 n/n_2)$, where $n$ indicates the $n$th patient in the Dose-ranging stage. After the stage concludes, to obtain an updated acceptable set of doses, we refit the TITE-CRM using all gathered toxicity data with the updated posterior probability of toxicity from the Dose-ranging stage, as the prior toxicity skeleton. To select the final OBD for each selected covariate pattern, we use cutoff values of $\epsilon = 0.85$ and $\Psi_{\text{OBD}} = 0.35$, guided by the choice made in \citet{guoBayesianPhaseII2023} and calibration across a wide range of scenarios. For the final model fit for both efficacy and toxicity, we assumed all patients have been fully observed.

For each futility assessment, to find influential patient regression covariates, we set $\Psi_F$ to $0.65$. Thus, in order to remove a subgroup, stronger criteria had to be met when identifying the patient covariate. In the first check (Assessment 1), a sufficient number of patients had been assigned if at least 6 had been treated at $\text{MTD}^*$. In the second check (Assessment 2), 10 patients in the subgroup must have been assigned across the range of acceptable doses. This ensured sufficient information to stop enrolling the subgroup. To determine futility, a cutoff of $C_F = 0.40$ and a threshold of $\lambda = 0.05$ was used, assuming an efficacy of 0.40 or lower is not acceptable for the therapy of interest.

\subsection{Performance Measures}

We focus on three areas of the design to assess performance: the use of futility assessments to identify the target population, the identification of true OBD, and the use of variable selection to identify the correct subgroups within the target population. Firstly, we assess the identification of the target population, which is determined by futility assessments to remove subpopulations that do not benefit from therapy. We measure the proportion of time the correct and incorrect target population is identified and the stage of the trial at which the target population is identified: Assessment 1, after the Dose-escalation stage, Assessment 2, before the Optimization phase, or Assessment 3, at the end of the trial when the OBD is estimated to not exist for a particular subgroup. The earlier a futile subgroup is identified, the fewer patients in this group are exposed to the therapy. We also include the proportion of time the trial stopped because the therapy was found to be futile for all subgroups. Correct identification of the target population occurs when all true unresponsive subpopulations are identified by the design when the target population is a subset of the patient population, or when no subgroup is identified as futile when the target population is the entire patient population. Incorrect identification of the target population occurs when the true futile subpopulations are not identified or when the incorrect subpopulations are found as unresponsive. Secondly, we measure the percent of correct OBD selection (PCS) for the target population, which will be stratified by each true subpopulation within the target population.
Thirdly, we are interested in the identification of the correct subpopulations in the target population, which is learned through variable selection. Specifically, we measure the true positive rate (TPR) and false positive rate (FPR) for covariate selection, meaning we assess the levels within each patient characteristic. We define TPR as the average number of influential covariates that differentiate the OBD selected in the final model divided by the total number of influential covariates that differentiate the OBD. The FPR is defined as the average number of non-influential covariates selected in the final model divided by the total number of non-influential covariates. 
Note that if only one level of a patient characteristic with more than two levels differentiates the OBD, even when other levels impact the probability of efficacy, as in Scenarios 1, 5, and 6, these levels are considered non-influential and contribute to the FPR.  
Finally, in addition to the three areas of performance, we also assess the impact of the PK-adjustment in the dose-optimization design and the importance of the reference level choice in the variable selection method through a sensitivity analysis.

We compare the proposed design to the proposed design without accounting for patient heterogeneity, the naive method, and the proposed design without accounting for PK. For the naive method, we use the same two stages of our design but do not include patient covariates in the efficacy model so there is no variable selection step and the design follows as if no covariates are identified as influential. The naive method also does not use futility assessments to identify the target population since no patient characteristics are incorporated. Therefore, we assume that after the Dose-escalation stage, when the first futility check would occur, the target population is known or correctly estimated from the empirical evidence gathered from $n_1$ patients. Traditionally, before dose-expansion cohorts are added, the target population is determined from the empirical evidence acquired in the initial dose-escalation trial and any prior information. Given the limited sample size, the information will not always be reflective of the true target population in reality. The method without PK-adjustment is carried out similarly to the proposed design except the acceptable dose set, $\mathcal{A}_d$, is not adjusted.

We conduct a sensitivity analysis for the variable selection method when an alternative reference level is used. We assess the performance of the method when the reference level is influential in identifying OBD or in identifying the target population. Specifically, in the first case, we rearrange the levels for the `gene type’ category to change the reference level to NTRK 1/2/3. We assess how results changed for Scenarios 5 and 6, since this category is influential in OBD. For the second case, we rearrange the levels in the `molecular alteration type’ category so that `gene fusion’ is the reference level and assess how the reordering impacts the performance in Scenarios 1, 2, 5, and 6 since in these scenarios the target population comprises patients with a gene fusion.

\subsection{Simulation Results}
\label{s:results}

\begin{figure}
\centering
\includegraphics[width=.8\textwidth]{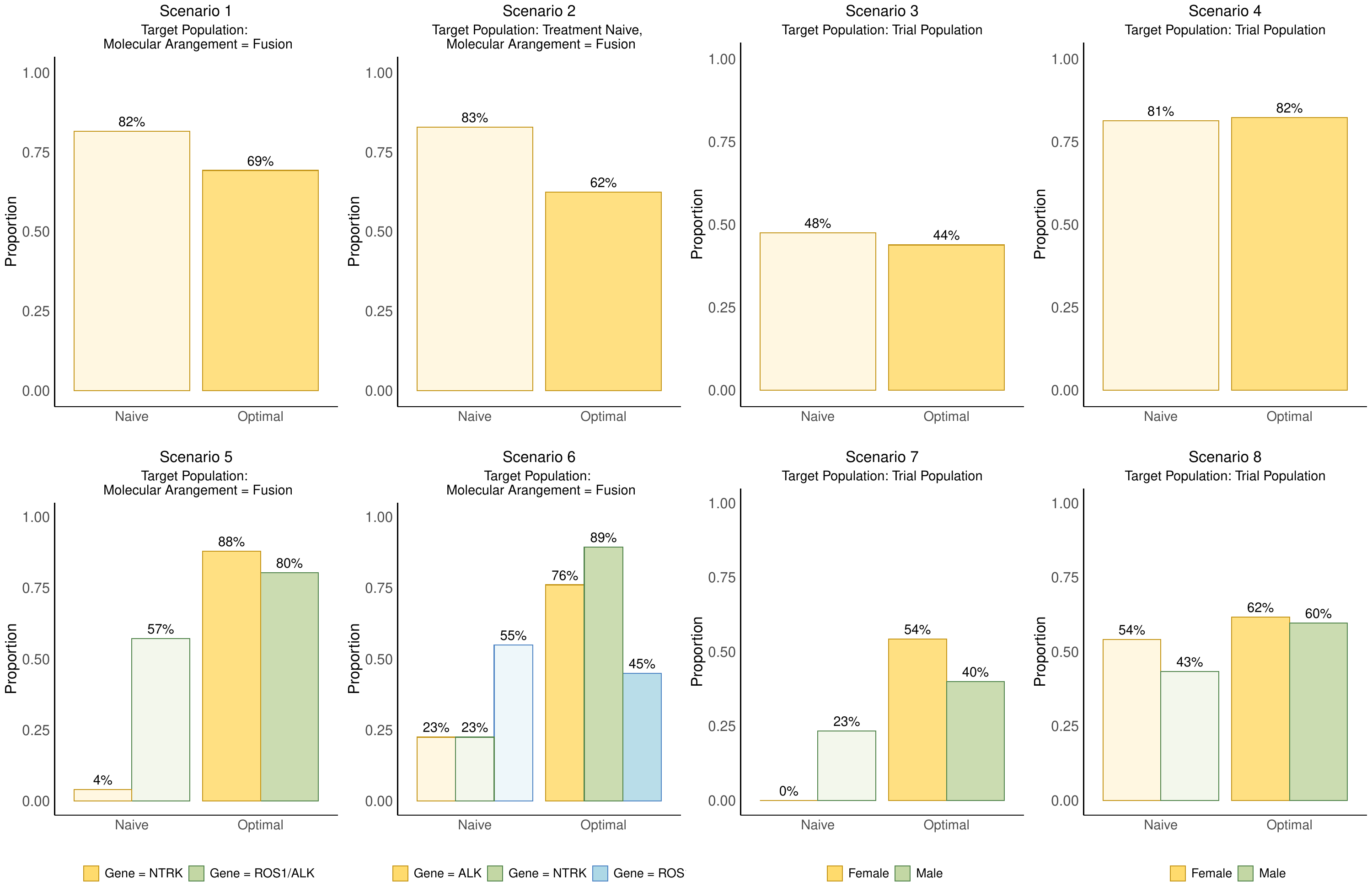}
\caption{Probability of Correct Selection of OBD (PCS) for $N_{\max} = 60$. OBD, Optimal Biological Dose.}
\label{pcs_figure}
\end{figure}

\begin{figure}
\centering
\includegraphics[width=.8\textwidth]{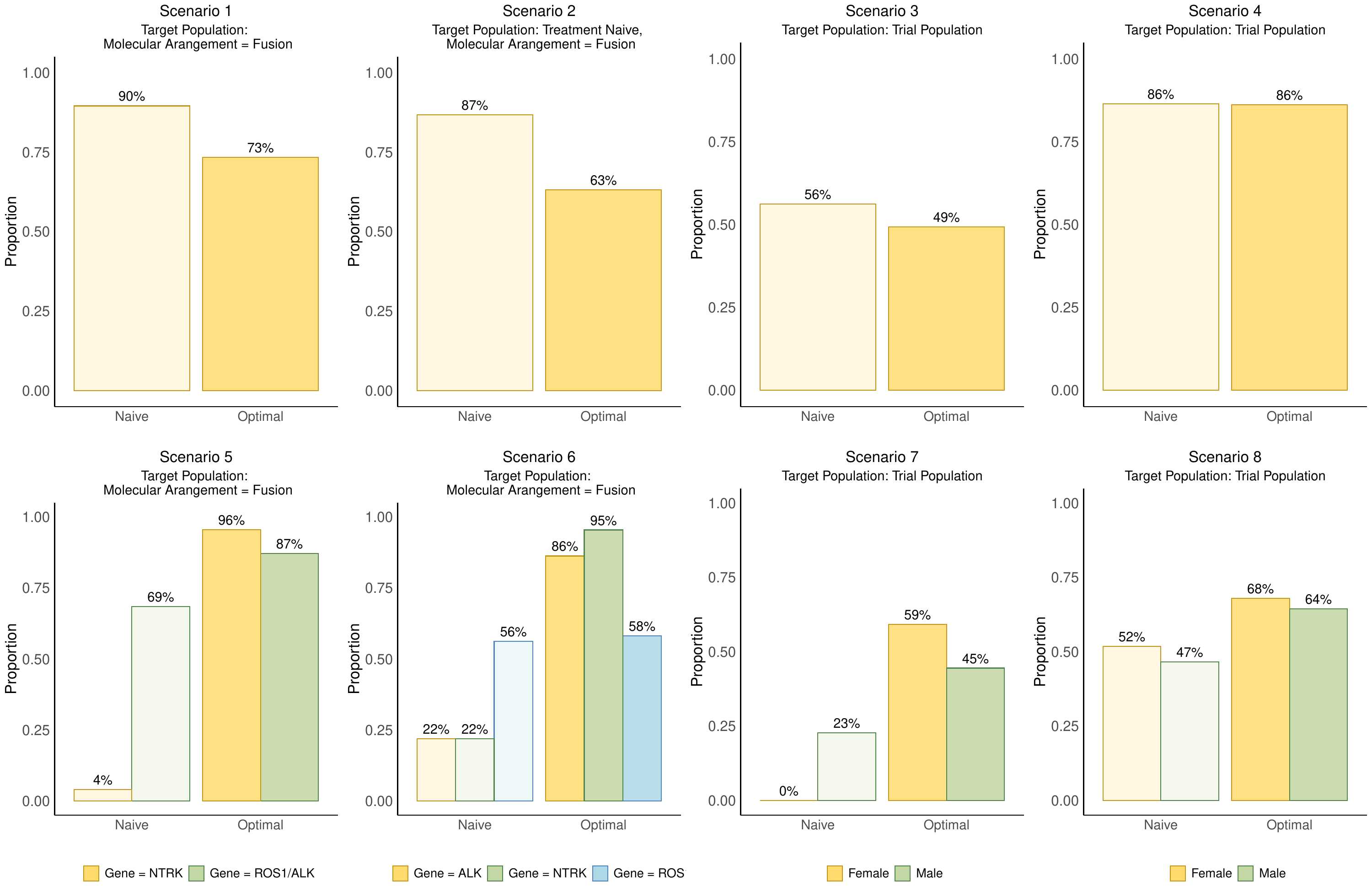}
\caption{Probability of Correct Selection of OBD (PCS) for $N_{\max} = 80$. OBD, Optimal Biological Dose.}
\label{pcs80_figure}
\end{figure}

The results of target population identification are shown in Table 2, giving the proportion of correct identification and the proportion of time the correct population was identified across futility assessments, and the proportion of incorrect identification. The percent of correct OBD selection (PCS) for the target population across true subpopulations is shown in Figure~\ref{pcs_figure} for $N_{\max} = 60$ and in Figure~\ref{pcs80_figure} 
for $N_{\max} = 80$. Lastly, the TPR and FPR for variable selection are given in Table 3 
for $N_{\max} = 60$ and $N_{\max} = 80$. 

\textit{Identification of target population:} 
Table 2 
gives the proportion of time a target population is identified when $N_{\max} = 60$. In addition to obtaining the proportion of correct target population identification, we also identify the stage at which it is identified for scenarios in which futile subpopulations exist (1, 2, 5, and 6). For incorrect target population identification, we differentiate between when no subgroup is eliminated and when the incorrect subgroup is eliminated, as the latter is less favorable. In Scenarios 1, 5, and 6, only one covariate differentiates the patient population between those who respond and those who do not. In these scenarios, the correct identification of the target population is around 90\% and it is most often identified in the second futility assessment, after $42$ ($n_1+\frac{n_2}{2}$) patients have been assigned a dose. The incorrect target population is identified less than 10\% of the time. In Scenario 2, two covariates determine the target population and must be identified. We find that both are identified 31\% of the time, most often during the second futility assessment and at the end of the trial, and at least one of the covariates is identified 84\% of the time. Thus, the incorrect target population is identified in 56\% of the trials since both covariates need to be identified. However, an incorrect futility subpopulation is only identified 6\% of the time, demonstrating that the rate of finding a subgroup not responsive when the subgroup is actually responsive is well-controlled. 
For $N_{\max} = 80$, the performance slightly improves; the proportion of correct target population increases for Scenario 2 to 40\% and increases by about 3\% for Scenarios 1, 5, and 6 while the identification of the incorrect target population remains controlled. Lastly, in Scenarios 3, 4, 7, and 8, when the target population is the entire population, the design is unlikely to eliminate any subpopulation, only doing so about 2.5\% of the time, on average.

\textit{Probability of Correct Selection of OBD:}
The probability of correct selection of OBD (PCS) for $N_{\max} = 60$ and $N_{\max} = 80$ is displayed in Figure~\ref{pcs_figure} and Figure~\ref{pcs80_figure}, respectively. The `Naive' method of accounting for no covariates and assuming homogeneity is compared to the proposed `Optimal' method.
In Scenarios 1, 2, 3, and 4, because the target population is homogeneous with respect to the OBD, and we assume that the naive method is able to correctly identify the target population through empirical evidence, it has the upper hand since the covariates add noise in the Optimal method. However, the Optimal method performs very similarly to the naive method whenever the target population is the trial population, as in Scenarios 3 and 4. In Scenarios 1 and 2, where the target population is defined by one or two patient characteristics, the Optimal method has about a 15-20\% decrease in PCS compared to the naive method since there is more heterogeneity to identify. For example, in Scenario 2, the target population is defined by the patient characteristics indicating prior treatment and the type of molecular alteration, so it takes more patients to identify the target subpopulation, making it harder to home in on the correct dose. In this scenario, the Optimal design still exceeds a PCS of 60\% by $N_{\max} = 60$. We also note that in these scenarios where the target population is a subset of the trial population, it may not be as realistic to assume the naive method knows the target population, so the performance of the naive method may be overestimated. 
For example, if we assume that the naive method does not know the target population after the initial stage, the naive method obtains a PCS of 69\% instead of 83\%. 
For Scenarios 5 and 6, there is heterogeneity in both the trial population, as in Scenarios 1 and 2, and within the target population. In Scenario 5, the target population now has two optimal doses depending on whether the patient has a NTRK 1/2/3 gene fusion, or ROS1 or ALK gene fusion. Since the ROS1/ALK gene fusion population makes up about two thirds of the target population based on the prevalence of the covariates, the naive method, which always assigns only one dose, favors that subgroup 57\% of the time, but since the OBDs are two doses apart, for almost the rest of the time, the naive method recommends the dose in-between, thus incorrectly dosing the entire target population. On the other hand, even though the NTRK 1/2/3 subpopulation only makes up about a third of the target population, the Optimal method identifies the OBD for both subpopulations over 80\% of the time.  In Scenario 6, we now have three subpopulations with different optimal doses, each one dose level apart from another. By recommending only one dose, the naive method always incorrectly doses about two thirds of the target population. The Optimal method identifies the OBDs at the boundary dose levels in the acceptable set (OBD 1 and OBD 3) at a high rate for both sample sizes but identifies the correct dose for the subpopulation with a ROS1 gene fusion less, with a PCS of 45\% for $N_{\max} = 60$ and 58\% for $N_{\max} = 80$. However, this subgroup PCS is still close to the best subgroup PCS for the naive method in this scenario, demonstrating the advantage of accounting for heterogeneity even in more complex scenarios.
In Scenarios 7 and 8, the target population is the entire population and the binary variable, gender, differentiates the OBD. In Scenario 7, the OBDs are two doses apart, so the naive method does very poorly since it tends to recommend the middle dose which is suboptimal for both subpopulations. While the Optimal method outperforms the naive method, it does not identify the boundary dose for the male subpopulation, dose level four, as well as dose level two, the OBD for the female subpopulation. Since the estimation favors a plateau relationship, it is split between recommending dose three (50\%) and four (40\%) for the male subpopulation. Additionally, since the toxicity skeleton identifies dose level 3 as the MTD, but all four doses are safe in this scenario, we find that the set of dose levels explored in the Dose-ranging stage did not include dose level 4 about one third of the time. While toxicity is evaluated again before OBD estimation and all four doses were defined as acceptable 87\% of the time, there might have not been enough exploration of dose level four to learn that it exhibits substantially higher efficacy than dose level three for that subpopulation. Lastly, in Scenario 8, the OBD is one dose apart and the PCS for the Optimal method reaches 60\% by $N_{\max} = 60$ for both subgroups while the PCS for the naive method stays around 50\% for both subgroups, for $N_{\max} = 60$ and $N_{\max} = 80$, since it alternates between recommending the higher and lower doses for the entire population.

\textit{Bayesian Variable Selection:}
The TPR and FPR for $N_{\max} = 60$ and $N_{\max} = 80$ across scenarios are shown in Table 3. 
Across all scenarios and total sample sizes, the FPR is controlled at or below 15\%. We note that in Scenarios 1, 2, and 5, there are one or two covariates that affect the probability of efficacy even though they do not differentiate the OBD, and therefore these covariates contribute to the FPR. When this is not the case (Scenarios 3, 4, 6, 7, and 8), the FPR is controlled below 5\%. For the scenarios in which patient characteristics do affect the OBD, the TPR increases as heterogeneity in OBD grows. For Scenarios 5, 6, and 7, where there are more than two doses to explore, the TPR is at least 76\% for $N_{\max} = 60$. For Scenario 8, when only two doses are safe, there is less exploration and thus less information about heterogeneity across doses since the response rates only substantially differ in one of the two doses. In this scenario, the TPR is only 37\% for $N_{\max} = 60$, however the FPR is controlled at 3\%, indicating that the design resorts to the default approach of assuming patient homogeneity rather than incorrectly identifying another patient characteristic. When $N_{\max} = 80$, the TPR increases by about 10\% on average, as more information can be learned from more patients.

We find that the PK-adjustment does not have an effect on the PCS compared to our proposed design without PK-adjustment. Most scenarios have about a 1\% increase in PCS with the adjustment, except for Scenario 2, where the adjustment decreases the PCS by about 2.5\%. In general, the PK-adjustment adds a conservative layer of safety before the Dose-ranging stage to mitigate exploring overly toxic doses. Specifically, the proportion of time the true set of acceptable doses, $\mathcal{A}_d^*$ with PK-adjustment and $\mathcal{A}_d$ with no PK-adjustment, are chosen only differs by 1.5\% in the two designs. However, when the correct set is not chosen, the proportion that the set of acceptable doses includes an overly toxic dose is about 9\% lower for the design with PK-adjustment and the proportion that the set of acceptable doses misses the true maximum dose in the set, is about 11\% higher for the design with PK-adjustment. Therefore, the design with PK-adjustment tends to be more conservative with regard to safety, choosing a lower upper threshold for toxicity. However, since toxicity is evaluated again before OBD estimation, the PCS is not substantially different. Since one aim of a dose-finding design is to minimize the number of patients assigned to an unsafe dose during the trial, if prior knowledge about PK exists, the PK-adjustment will ensure more safety during the trial without compromising overall performance in OBD estimation. 

Lastly, from the sensitivity analysis on reference levels, we found that the specification of the reference level is important since if it determines patient heterogeneity, the results of the trial may be affected. Firstly, when the OBD depends on the reference level, as in Scenarios 5 and 6 after reordering, both non-reference level categories of gene type need to be identified and thus the PCS and variable selection is affected, with an average decrease in PCS of 17\% and TPR decrease by 9\% for Scenario 5 and 31\% for Scenario 6. The identification of the target population is also affected, with a decrease of correct selection by about 10\% for Scenarios 5 and 6. Secondly, when the target population depends on the reference level, as in Scenarios 1, 5, and 6 after reordering, the correct identification of the target population is majorly affected. The target population is identified about 40\% less of the time since both non-reference levels need to be identified. The PCS was also affected, decreasing on average by about 15\% when heterogeneity in OBD exists, as in Scenarios 5 and 6. Furthermore, when the target population is not identified, there is more heterogeneity observed among the assumed patient population, which led to about a 25\% decrease in TPR for variable selection. Therefore, the reference level choice is important and although prior understanding of the effects of certain characteristics are not required, when a level of a category makes up a large proportion of the population, as was the case for molecular alteration type of `gene fusion', it should not be chosen as the reference level.

\section{Discussion}
\label{discuss}

The proposed design addresses several key considerations of Project Optimus, including randomization to more than one dose, plateauing or monotone efficacy profiles, PK incorporation, target subpopulations, and delayed onset of trial outcomes. While continuing to ensure safety, dose-finding in the new paradigm must adapt from sequential dose-assignment to randomized exploration of a range of safe doses to address the `less is more’ rationale of Project Optimus. Additionally, even though efficacy will be evaluated in the first phase of drug development, the target population may not be solidified during enrollment. Therefore, the identification of the target populations during the trial and the ability to adjust as more information is learned will improve the efficiency of dose-finding. Our design addresses these intricacies through a model-based approach and can incorporate prior information under a Bayesian framework. The proposed dose-optimization design is recommended when heterogeneity in efficacy is suspected in the patient population, but the sources of such heterogeneity, i.e., patient covariates accountable for the heterogeneity, are to be determined. The design can accommodate many potential predictive markers of efficacy and identify both the markers that affect whether the therapy is futile for some and the markers that affect the recommendation of OBD.  

The proposed design was applied to a trial evaluating a molecularly targeted agent for which the target population was not assumed, and more than one optimal dose could exist. We found that when one covariate differentiates the trial population that responds to the treatment, the design identifies the target population at a high rate and most often does so before all patients are assigned a dose, minimizing the enrollment of patients who are unlikely to respond. While the selection of the target population when it was defined by two covariates, was fully correct only about a third of the time, this case would also be difficult with empirical evidence, without a model-based approach, and can lead to a lot of uncertainty. We found that at least one covariate was identified most of the time, demonstrating that the modeling approach is still worthwhile and would not eliminate any subpopulations too early. A more refined target population could be identified in later phases. The approach of identifying the target population through modeling heterogeneity is novel and no other method has addressed this formally, apart from utilizing futility stopping rules for predefined subpopulations. We found that the second stage is needed to accurately identify the target population since after the initial dose-escalation stage, not enough heterogeneity will be observed.
Additionally, while the TPR was most favorable when the OBD differed by more than one dose, the design still chose the correct OBD most often and outperformed the naive approach in any scenario with heterogeneity. The design also showed control of the FPR, indicating the design defaults to the naive approach if a strong enough signal does not exist rather than choosing an incorrect characteristic. We also find better TPR when more doses are in the safe set and thus explored, highlighting the importance of having a defined acceptable dose set to work from and then randomizing across the dose set to explore heterogeneity.
Finally, the PCS performance shows that in the presence of homogeneity, the design does not do much worse than the naive approach, validating that in any case where heterogeneity is suspected but not well understood, the design is advantageous.


Limitations of the proposed dose-optimization design and the corresponding simulation study include not integrating certain trial information into the design, certain model assumptions, and the lack of comparison methods. The current design does not include a PD biomarker which can provide initial evidence of efficacy. Ideally, this would be incorporated when defining a set of doses after the dose-escalation study, where the PD endpoint and MTD could help identify the lower and upper bound, respectively, as done in \citet{guoDROIDDoserangingApproach}. However, in the presence of heterogeneity in efficacy and with a limited number of patients in the initial stage, there could be too much uncertainty to justify eliminating a low dose early on. Additionally, while we adjusted the PK threshold that determined toxicity, we assumed that the threshold was always known. Furthermore, we have assumed the therapy is generally well tolerated and tolerated similarly across patients since many patient biomarkers contribute only to efficacy, however, some characteristics may also influence toxicity and specification of more than one safe set of doses may be desirable. We have also considered incorporating patient-reported outcomes (PRO), using the PRO-CRM, proposed by \citet{leeIncorporatingPatientReportedOutcomes2020}, instead of using the TITE-CRM in the dose-escalation phase since the patient perspective on toxicity is as an important aspect of the toxicity burden. We do not advise adjusting for the PRO-MTD similarly to our PK-incorporation since simulations found that only adjusting the MTD at the end instead of using patient and clinician defined toxicity throughout, lead to conservative estimations of the MTD \citep{leeIncorporatingPatientReportedOutcomes2020}. There are also limitations concerning the working model for efficacy. Firstly, the working model makes the strong assumption that all efficacy curves have the same plateau limit. Thus, even though the dose at which efficacy plateaus can be different, it assumes the same probability of efficacy is reached. While this is a restrictive model, we note that all simulation study scenarios had efficacy curves that plateaued at different probabilities and the design was still able to identify the different OBDs. Secondly, the working model does not account for non-monotone, unimodal efficacy profiles and we have not studied the design’s robustness for these curves. We recommend that this method is applied when a strictly monotone or plateau profile is suspected, as is the case for most cancer therapies. Lastly, in the simulation study, we do not compare our proposed method to any phase I/II designs that account for patient heterogeneity. This is because these methods are optimal for predefined subpopulations or a predefined number of subgroups that are heterogeneous, and a predefined target population. 

In conclusion, there is a major benefit to include patient characteristics and sufficiently explore and understand the therapy at an early phase. In the era of precision medicine, where we are learning about the many factors that can affect patient response and why a one-size-fits-all approach should not apply to all therapies, the earlier we can understand heterogeneity, the more effective a cancer therapy can be. 
Our method can be applied to settings where there is a fixed population and enrollment criteria do not change over the course of the trial, and to settings where the population may change during the trial due to changes in enrollment criteria as more information about the target population is gathered.
Most existing methods do not allow this flexibility, and the flexibility of our method ensures that helping expedite drug development does not come at the cost of requiring too many assumptions about the patient population. 
While including patient covariates and using a Bayesian model-based approach may require more patients and resources, ignoring the complexities that help explain the dose-toxicity and dose-response relationships may require repeating previous studies and analyses or discontinuing therapies that could be beneficial for a specific population that was not sufficiently explored.

\newpage

\bibliographystyle{rss}
\bibliography{references}


\begin{table}

\caption{Probability of Toxicity and Efficacy Across Scenarios}
\begin{threeparttable}
\begin{tabular}{p{.6cm}p{1.20cm}p{1cm}p{1.5cm}p{1.2cm}p{1.2cm}p{1.17cm}p{1.17cm}p{1.17cm}p{1.17cm}}
  \toprule
  & Subgroup & Gene & Mutation & Prior Tx & Gender & D1  & D2  & D3  & D4 \\ 
S1 &  &  &  & &  & (0.05) &  (0.12) &  (0.25) & (0.38) \\ 
    & 1 & NTRK & Fusion &\small{NA}&\small{NA}& 0.50 & 0.70 & \textbf{0.95} & 0.95 \\ 
     & 1 & ROS1 & Fusion &\small{NA}&\small{NA}& 0.25 & 0.55 & \textbf{0.86} & 0.86 \\ 
      & 1 & ALK & Fusion &\small{NA}&\small{NA}& 0.10 & 0.35 & \textbf{0.57} & 0.57 \\ 
       & NA &\small{NA}& Other &\small{NA}&\small{NA}& 0.05 & 0.05 & 0.05 & 0.05 \\ 
S2 &  &  &  &  &  & (0.05) & (0.12) & (0.25) & (0.38) \\ 
                & 1 & NTRK & Fusion & No &\small{NA}& 0.50 & 0.70 & \textbf{0.95} & 0.95 \\ 
                 & 1 & ROS1 & Fusion & No &\small{NA}& 0.25 & 0.55 & \textbf{0.86} & 0.86 \\ 
                  & 1 & ALK & Fusion & No &\small{NA}& 0.10 & 0.35 & \textbf{0.57} & 0.57 \\ 
                   & NA &\small{NA}& Other & Yes &\small{NA}& 0.05 & 0.05 & 0.05 & 0.05 \\ 
S3 &  &  &  &  &  & (0.02) & (0.06) & (0.14) & (0.24) \\ 
                        & 1 &\small{NA}&\small{NA}&\small{NA}&\small{NA}& 0.5 & \textbf{0.7} & 0.7 & 0.7 \\ 
S4 &  &  &  &  &  & (0.05) & (0.12) & (0.25) & (0.38) \\ 
                         & 1 &\small{NA}&\small{NA}&\small{NA}&\small{NA}& 0.25 & 0.5 & \textbf{0.7} & 0.7 \\ 
  S5 &  &  &  &  &  & (0.05) & (0.12) & (0.25) & (0.38) \\ 
        & 1 & NTRK & Fusion &\small{NA}&\small{NA}& \textbf{0.95} &  0.95 & 0.95 & 0.95 \\ 
         & 2 & ROS1 & Fusion &\small{NA}&\small{NA}& 0.25 & 0.55 & \textbf{0.86} & 0.86 \\ 
          & 2 & ALK & Fusion &\small{NA}&\small{NA}& 0.10 & 0.35 & \textbf{0.57} & 0.57 \\ 
           & NA &\small{NA}& Other &\small{NA}&\small{NA}& 0.05 & 0.05 & 0.05 & 0.05 \\ 
  S6 &  &  &  &  &  & (0.05) & (0.12) & (0.25) & (0.38) \\ 
            & 1 & NTRK& Fusion &\small{NA}&\small{NA}& \textbf{0.95} & 0.95 & 0.95 & 0.95 \\ 
             & 2 & ROS1 & Fusion &\small{NA}&\small{NA}& 0.55 & \textbf{0.86} & 0.86 & 0.86 \\ 
              & 3 & ALK & Fusion &\small{NA}&\small{NA}& 0.2 & 0.35 & \textbf{0.57} & 0.57 \\ 
               & NA &\small{NA}& Other &\small{NA}&\small{NA}& 0.05 & 0.05 & 0.05 & 0.05 \\ 
   
  S7 &  &  &  &  &  & (0.02) & (0.06) & (0.14) & (0.24) \\ 
                    & 1 &\small{NA}&\small{NA}&\small{NA}& Male & 0.15 & 0.25 & 0.35 & \textbf{0.6} \\ 
                     & 2 &\small{NA}&\small{NA}&\small{NA}& Female & 0.55 & \textbf{0.8} & 0.8 & 0.8 \\ 
  S8 &  &  &  &  &  & (0.12) & (0.23) & (0.41) & (0.56) \\ 
                   & 1 &\small{NA}&\small{NA}&\small{NA}& Male & 0.35 & \textbf{0.6} & 0.6 & 0.6 \\ 
                       & 2 &\small{NA}&\small{NA}&\small{NA}& Female & \textbf{0.7} & 0.7 & 0.7 & 0.7 \\ 

   \bottomrule
\end{tabular}
\begin{tablenotes}
\item[] Notes: OBD in bold. Probability of toxicity in parentheses. Subgroup is determined by OBD. NA under Subgroup indicates not in target population, otherwise NA indicates non-influential patient characteristic. Tx Treatment.
\end{tablenotes}

\end{threeparttable}

\label{scenario_tblp}
\end{table}

\begin{table}
\caption{Proportion of Identification of Target Population for $N_{\max} = 60$}
\begin{threeparttable}
    \begin{tabular}{lllll}
    \toprule
 Scenario & Correct Target  & Incorrect Target  & Partially Correct & Trial Ended \\ 
 &  Population  & Population &  Target & Early \\
  & (Assessment 1, 2, 3)  & (Incorrect subgroup  & Population &  \\ 
  
&   & identified) & & \\ 
1 & 0.91  (0.30, 0.58, 0.02) & 0.07 (0.06) & \small{NA}& 0.03 \\ 
  2 & 0.31 (0.15, 0.22, 0.21) & 0.56 (0.06) & 0.84 & 0.13\\
3 & 1 & 0 & \small{NA} & 0.00\\ 
  4 & .99 & 0.01 & \small{NA}&0.00\\ 
  5 & 0.90 (0.31, 0.57, 0.02) & 0.08 (0.07) & \small{NA}& 0.03\\ 
  6 & 0.91 (0.34, 0.56, 0.02) & 0.08 (0.05) & \small{NA}& 0.01 \\ 
  7 & 0.93 & 0.06 & \small{NA} & 0.01\\ 
  8 & 0.97 & 0.03 & \small{NA}& 0.03\\ 
   \bottomrule
\end{tabular}
\begin{tablenotes}
\item[] Notes: Assessment 1 and 2 are the first and second futility assessments, after the dose-escalation stage and before the Optimization phase, respectively. Assessment 3 occurs at the end of the trial if OBD is estimated to not exist. Correct identification in Scenarios 1, 2, 5, and 6, imply true futile subgroups were identified. Correct identification in Scenarios 3, 4, 7, and 8, imply no subgroups were identified as not responsive to the therapy. 
Correct elimination for Scenarios 1, 5, and 6 indicates that patients with a gene fusion were identified as the target population. Correct elimination for Scenario 2 indicates that patients with a gene fusion and no prior treatment were identified as the target population. Partially correct elimination for Scenario 2 indicates that at least one subgroup was eliminated. NA = Not applicable. 
\end{tablenotes}
\label{futilty_n60}
\end{threeparttable}
\end{table}

\begin{table}
\caption{True Positive Rate (TPR) and False Positive Rate (FPR) of Covariate Selection for $N_{\max} = 60, 80$}
  \centering 
  \begin{threeparttable}
     \begin{tabular}{llll}
   \toprule
  $N_{\max}$ & Scenario & TPR & FPR \\  
  60 & 1 & \small{NA} & 11 \\ 
 & 2 & \small{NA}  & 7 \\
  & 3 &  \small{NA} & 4 \\ 
 & 4 & \small{NA}  & 2 \\ 
 & 5 &  85 & 8 \\ 
  &6 &  76 & 2 \\ 
 & 7 &  82 & 3 \\ 
 & 8 &  37 & 3 \\ 

 80 & 1 & \small{NA} & 15 \\ 
 & 2 & \small{NA} & 7 \\ 
 & 3 & \small{NA} &3 \\ 
&  4 & \small{NA} & 2 \\ 
& 5 & 94 & 10 \\ 
 & 6 & 86 & 2 \\ 
 & 7 & 92 & 3 \\ 
&  8 & 49 & 3 \\ 
   \bottomrule
   \end{tabular}
   \begin{tablenotes}
   \item[] Notes: Scenarios 1 through 4 have no patient characteristics associated with OBD. NA = Not applicable.
\end{tablenotes}
   \label{bvs_table}
\end{threeparttable}
\end{table}

\end{document}